\documentclass[letterpaper,journal]{IEEEtran}
\usepackage{cite}
\usepackage{amsmath,amssymb,amsfonts}
\usepackage{graphicx}
\usepackage[table]{xcolor} 
\usepackage{textcomp}
\usepackage{stfloats}  
\usepackage{multirow}  
\usepackage{hyperref}
\hypersetup{
    colorlinks=true, 
    linkcolor=blue, 
    pdfborder={0 0 0}, 
    filecolor=blue, 
    urlcolor=black,
    citecolor=blue
}

\usepackage{tikz}
\newcommand\noticetext{\footnotesize This paper has been accepted for publication by IEEE. C. Joglekar, C. Eze, D. Xiang and A. Monti, "A Hybrid Intrusion Detection System for Electric Vehicle Charging Infrastructure," in IEEE Internet of Things Journal, doi: 10.1109/JIOT.2026.3725487.
Upon publication, it will be made available under a Creative Commons Attribution 4.0 International License.
}
\newcommand\submittednotice{%
  \begin{tikzpicture}[remember picture,overlay]
    \node[anchor=south,yshift=10pt] at (current page.south)
      {\fbox{\parbox{\dimexpr\textwidth-\fboxsep-\fboxrule\relax}{\noticetext}}};
  \end{tikzpicture}}
  
\begin{document}

\title{A Hybrid Intrusion Detection System for Electric Vehicle Charging Infrastructure}

\newcommand{\orcidlink}[1]{%
  \href{https://orcid.org/#1}{\includegraphics[height=0.9em]{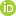}}%
}

\author{Charukeshi Joglekar\orcidlink{0000-0003-2642-8868},~\IEEEmembership{Member,~IEEE,}
        Chijioke Eze\orcidlink{0000-0002-8545-0160},~\IEEEmembership{Graduate Student Member,~IEEE,}
        Danni Xiang\orcidlink{0009-0002-4891-9270},
        and Antonello Monti\orcidlink{0000-0003-1914-9801},~\IEEEmembership{Senior Member,~IEEE}
\thanks{This work was supported by the project End-to-end Cybersecurity to NEMO meta-OS (CyberNEMO), funded by the European Union’s HORIZON Innovation Actions under Grant Agreement No.: 101168182.~\textit{(Charukeshi Joglekar and Chijioke Eze contributed equally to this work.)  (Corresponding author: Charukeshi Joglekar.)}}
\thanks{Charukeshi Joglekar and Antonello Monti are with the Fraunhofer Institute for Applied Information Technology, 52068 Aachen, Germany, and also with the Institute for Automation of Complex Power Systems, RWTH Aachen University, 52074 Aachen, Germany (e-mail:~charukeshi.mayuresh.joglekar@fit.fraunhofer.de, antonello.monti@eonerc.rwth-aachen.de).}
\thanks{Chijioke Eze and Danni Xiang are with the Institute for Automation of Complex Power Systems, RWTH Aachen University, 52074 Aachen, Germany (e-mail:chijioke.eze@eonerc.rwth-aachen.de;~danni.xiang@rwth-aachen.de).}
\submittednotice
}

\maketitle

\begin{abstract}
The integration of Electric Vehicle Charging Stations (EVCSs) into the smart grid necessitates sophisticated digital infrastructure for their management and coordination, which expands the attack surface and makes both the power grid and EVCSs vulnerable to cyberattacks. This research addresses critical gaps in existing EVCS Intrusion Detection Systems (IDS) by proposing
a Hybrid IDS that integrates attack detection on both the cyber and physical layers of the EV charging ecosystem. The proposed Hybrid IDS utilizes a dual-layer integration method, which combines network-based IDS (NIDS) and host-based IDS (HIDS). This approach facilitates comprehensive monitoring of both network traffic through the NIDS and host-level activities via the HIDS, which analyzes host events and power consumption data, effectively addressing the unique challenges posed by the interconnected nature of EV charging ecosystems. Utilizing the recent CICEVSE2024 dataset, the IDS presented in this work performs multiclass classification across various attack types, including False Data Injection Attacks (FDIAs), reconnaissance, denial of service, backdoor, and cryptojacking attacks. Experimental results demonstrate that our approach achieves excellent detection accuracy, with the NIDS component reaching 99.99\% accuracy for network-based attacks and the HIDS component achieving 83.47\% overall accuracy for host-based intrusion detection. In particular, the HIDS achieves F1-scores of 0.96 or higher for FDIA, backdoor, and cryptojacking attacks. This dual-layer detection provides more comprehensive attack coverage than the single-source detection approaches previously presented in the literature.
\end{abstract}

\begin{IEEEkeywords}
Electric vehicle charging stations, cybersecurity, intrusion detection, machine learning, network security, host security, power consumption
\end{IEEEkeywords}

\section{Introduction}
\IEEEPARstart{T}{he} global transition toward sustainable transportation has accelerated the deployment of Electric Vehicle Charging Stations (EVCSs).
At the end of 2024, there were approximately 5.2 million public EVCSs worldwide, with the number continuing to grow rapidly \cite{iea2025chargingpoints}. 
As the integration of Electric Vehicle Charging Stations (EVCSs) into the grid increases, it requires sophisticated digital infrastructures as well as communication between different actors to ensure coordinated charging from both the grid and the consumer perspective.

\begin{figure}[!t]
    \centering
    \includegraphics[width=\linewidth]{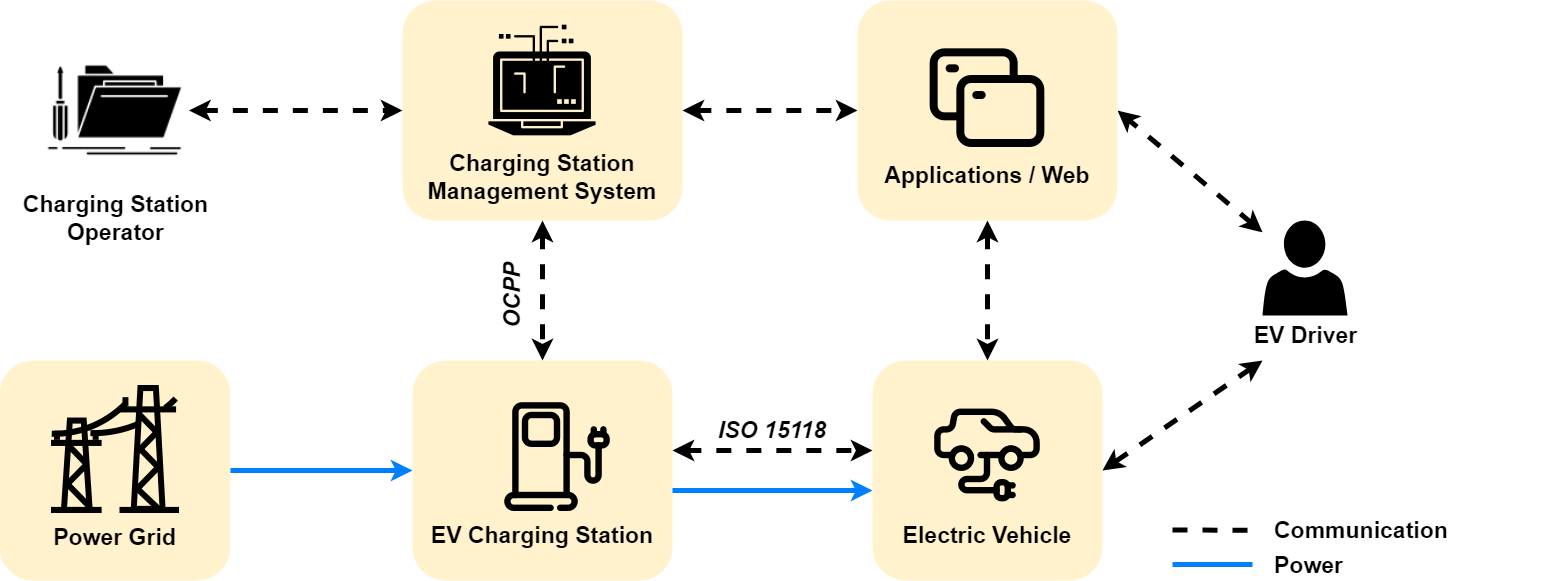}
    \caption{Overview of the EV charging ecosystem}
    \label{fig:evcs_ecosys}
\end{figure}

As shown in Fig. \ref{fig:evcs_ecosys}, the EV charging ecosystem is a complex system composed of a number of components managed by various actors. The components include: (1) Electric Vehicles (EVs); (2) EVCSs, which deliver power to EVs; (3) the Charging Station Management System (CSMS), which provides functionalities for managing the operation of all connected EVCSs; (4) applications/web interfaces, which allow users to locate, reserve, and pay for charging services; and (5) the power grid, which supplies electricity to the entire system.

The two primary actors in this ecosystem are: Charging Point Operators (CPOs), who own and operate the charging stations; and EV drivers, who use the charging services. These components and actors interact through various communication protocols. The communication between the CSMS, which is managed by the CPO, and the EVCS uses the Open Charge Point Protocol (OCPP), whereas the communication between EVCSs and EVs uses the ISO 15118 protocol. The ISO 15118 protocol supports V2G communication and relevant functionalities. Other actors such as Electromobility Roaming Service Providers, Distribution System Operators (DSOs) also communicate with the aforementioned actors for the control and coordination of EV charging to ensure grid reliability while fulfilling EV consumer requirements. 

The EV charging ecosystem is thus a cyber-physical system where digital communications control physical power flows. The interconnected nature of the system creates multiple potential attack surfaces that malicious actors can exploit to disrupt EVCS as well as grid operation \cite{Acharya2024_ev_attack_grid} (see Section \ref{sec:threats}). As Sharma et al. \cite{sharma2025artificial} point out, sophisticated intrusion detection mechanisms are required that facilitate attack detection at multiple layers of the cyber-physical ecosystem. 

Intrusion Detection Systems (IDSs) proposed in literature for EVCSs have generally been single-source detection approaches, i.e., either Network-based IDS (NIDS) or Host-based IDS (HIDS). Each of these has notable limitations regarding detection comprehensiveness and dataset relevance. Machine Learning (ML) classification algorithms, primarily supervised learning, have been widely used in prior studies \cite{MLIDSEVCS, DLIDSEVCS, HIDSEVCSessions}, where labeled data is required for training and prediction. These algorithms are validated for efficient and accurate attack detection and are commonly used in IDS development. 

Prior studies on NIDSs \cite{DLIDSEVCS, MLIDSEVCS} have typically relied on attack datasets containing both benign and malicious traffic. However, these datasets were derived from non-EVCS environments, limiting their applicability to EVCS-specific network environments. Conversely, existing HIDSs \cite{HIDSEVCSessions, HMMHIDSevent} have used datasets specific to EVCSs, but these contain only benign data, necessitating the use of simulated attack data that may not fully reflect real-world attack patterns.  

Recent advancements in deep learning, as seen in the work of Li et al. \cite{li2025multi} with multi-view graph contrastive learning and Benfarhat et al. \cite{benfarhat2025advanced} with temporal convolutional networks, have shown promising results in EVCS intrusion detection. However, these approaches typically focus on a single data source, not the comprehensive view that a hybrid approach can provide.

To address these limitations, this paper proposes a Hybrid IDS for the EV charging ecosystem based on the recently released CICEVSE2024 dataset \cite{CICEVSE2024}, which encompasses network traffic data, power consumption data, and host activities of EVCSs under both benign and attack scenarios. The proposed Hybrid IDS comprises a NIDS deployed at the CSMS in the cyber layer and HIDSs placed at individual EVCSs. The NIDS monitors overall network traffic, detecting network-based attacks that may involve multiple EVCSs, while HIDSs at each EVCS provide localized detection by monitoring log events and power consumption, primarily focused on identifying host-based attacks.

The main contributions of this paper are as follows. (1) We propose a dual-layered Hybrid IDS architecture that aligns with the cyber-physical structure of the EV charging ecosystem, providing comprehensive security monitoring. (2) We implement and evaluate machine learning-based multiclass classification for detecting and categorizing network-based and host-based attacks, including backdoor, cryptojacking, FDIAs, and different variants of reconnaissance and DoS attacks. (3) We develop and validate the proposed IDS using the CICEVSE2024 dataset, which contains real-world attack data specific to EVCSs rather than simulated attacks or general network or IoT related attacks. (4) We empirically demonstrate that the data preprocessing and feature engineering pipeline proposed in this work results in improved attack detection performance, i.e., higher detection accuracy and better attack classification compared to the results reported on the same dataset in literature.

The remainder of this paper is organized as follows: Section II discusses EVCS threat landscape, and presents related work on IDS for EVCSs. Section III presents the conceptual architecture of our proposed Hybrid IDS and describes the CICEVSE2024 dataset. Section IV details the implementation methodology and experimentation setup. Section V presents the evaluation results and comparative analysis. Finally, Section VI concludes the paper and suggests future research directions.

\section{Background and Related Work}

\subsection{Threat Landscape for EVCS} \label{sec:threats}
The EV charging ecosystem faces numerous security threats. Benfarhat et al. \cite{benfarhat2025advanced} identify up to 16 different attack types against EV charging systems, including both traditional network attacks and emerging threats specific to charging infrastructure. Considering the complex nature of the EV charging ecosystem, the Intrusion Kill Chain \cite{hutchins2011intelligence} introduced by Hutchins et al. provides a structured framework to describe the stages attackers go through to execute attacks on such a system. These are reconnaissance, weaponization, delivery, exploitation, installation, command and control, and actions on objectives \cite{hutchins2011intelligence}.
The first step in an attack is reconnaissance. In the context of EVCSs, reconnaissance attacks can be used by attackers to identify vulnerabilities in the communication between EVs, EVCSs, and the CSMS. These vulnerabilities can then be exploited to launch more sophisticated attacks on these components.

In the weaponization stage, an attacker can execute host-based attacks, such as backdoor attacks and cryptojacking, which directly target the EVCS hardware. Backdoor attacks can provide unauthorized access to attackers, while cryptojacking uses the computing resources of EVCSs for cryptocurrency mining without authorization. Particularly, backdoor attacks can be escalated further to orchestrate disruptive attacks that have adverse consequences for the individual EVCS as well as interconnected devices in the EV charging ecosystem. Therefore, the detection of host-based attacks at the level of individual EVCSs can reduce the likelihood of subsequent attacks.

In the context of attacks on the communication interfaces among EV charging ecosystem components, the work of Zhdanova et al. \cite{LPGatRisk} identifies two main classes of network attacks that pose threats to V2G communication. First, vulnerabilities in the ISO 15118 protocol can be exploited to conduct Denial of Service (DoS) attacks by flooding communication channels, as well as jamming attacks by transmitting jamming signals between the EVCS and EV. Second, attackers can exploit the vulnerabilities of the Open Charge Point Protocol (OCPP) to execute False Data Injection Attacks (FDIAs) by manipulating voltage, current, or power data, or to launch Man-in-the-Middle (MitM) attacks that intercept communication between the CPO and EVCS.

Sharma et al. \cite{sharma2025artificial} categorize EVCS threats into three key domains: grid-side attacks targeting the power grid infrastructure, communication-side attacks exploiting vulnerabilities in protocols like OCPP, and user-side attacks that target charging station access points and user interfaces. Their research highlights that positioning of EVCS at the intersection of energy and transportation infrastructure makes it particularly vulnerable to sophisticated attacks that could impact both domains simultaneously.

These attacks can have severe consequences, including power grid instability \cite{Acharya2024_ev_attack_grid}, financial losses for service providers \cite{Charu_2022}, and compromised availability of charging services for EV users \cite{Gumrukcu2024}. These impacts of cyberattacks on EVCSs emphasize the need for robust security measures that protect both cyber and physical layers of the ecosystem.

Further, Warraich and Morsi \cite{warraich2023early} specifically examine the threat landscape for fast charging stations, noting that the high power rates involved create additional risks. In particular, compromised charging sessions could lead to more immediate and severe impacts on grid stability, especially in microgrid environments where V2G operations are implemented.

Therefore, intrusion detection systems for the EV charging ecosystem must be able to classify network-based and host-based attacks to mitigate their impact on individual EVCSs as well as on the power grid.

\subsection{Existing IDS Approaches}

Existing research on IDSs for EVCSs can be broadly categorized into two approaches, which are discussed in Sections~\ref{sec:nids} and~\ref{sec:hids}.

\subsubsection{Network-based IDS (NIDS) for EVCS} \label{sec:nids}
The network layer of the EVCS is one of the key points that attackers use to orchestrate attacks against the system. This has attracted the attention of researchers within the EVCS security community. As a result, many detection approaches have been proposed. For example, ElKashlan et al. \cite{MLIDSEVCS} developed a machine learning-based NIDS for IoT EV charging stations. In their work, they propose a random forest classifier trained using features derived from the IoT23 dataset, which contains benign and malicious network traffic. Their approach achieved good accuracy but its reliance on a general IoT dataset \cite{garcia2020iot23} rather than EVCS-specific data implies it may not guarantee good performance in a typical EVCS environment.
Similarly, Basnet and Ali \cite{DLIDSEVCS} implemented a deep learning-based IDS for electric vehicle charging stations. They used the CICIDS2017 dataset, containing benign and various attack traffic patterns, and demonstrated that deep learning models could effectively detect network-based attacks. However, the dataset was not specific to EVCS environments. 

More recently, Benfarhat et al. \cite{benfarhat2025advanced} proposed an advanced Temporal Convolutional Network (TCN) framework specifically for intrusion detection in EVCSs. Their proposed MRG-ID-SA-TCN model integrates multi-receptive fields, gating mechanisms, iterative dilation, and self-attention to effectively capture temporal dependencies in network traffic. This approach achieved state-of-the-art results on the CICEVSE2024 dataset, with particularly strong performance in detecting denial-of-service and reconnaissance attacks. However, despite its good performance on multi-class attack detection for EVCS, again, the method leaves critical gaps in terms of real-world deployment considerations, including cross-dataset generalization, explainability, and continuous learning for evolving threats. In Jiang et al. \cite{Jiang_CICEVESE_KD}, the authors present a knowledge distillation enhanced semi-supervised Federated Learning (FL) framework for intrusion detection in EV charging networks. Although their model outperforms baseline methods from semi-supervised as well as federated learning, their proposed approach was only evaluated using the network traffic data from the CICEVSE2024 dataset; host-based attacks are not taken into consideration in their work.

Unlike these group of works, the present paper is not only validated on a realistic EVCS dataset to meet the unique nature of network based intrusions in EVCS environment but also includes detection of host based intrusions at the same time.   

\subsubsection{Host-based IDS (HIDS) for EVCS} \label{sec:hids}
Another important entry point for attacks against EVCS is the host devices installed within the charging infrastructure. An example is the backdoor attack, which can be exploited to compromise the charging infrastructure or connected systems, to steal data, or to cause damage to the EVCS. Many methods have been proposed in the literature to address the host intrusion attacks in EVCS. For instance, Kern et al. \cite{HIDSEVCSessions} implement a centralized HIDS to classify EV charging sessions as normal or anomalous. Their system combines regression-based charging behavior prediction with ensemble anomaly detection, incorporating classification-based and novelty-based detection methods. They used real-world EV charging session data from the ACN \cite{lee_acndata_2019} and ElaadNL \cite{elaadnl2019charging} datasets, which contain only benign charging data, to validate their proposed method. This raises questions about the applicability of their method in distinguishing different host-based attacks. In Girdhar et al. \cite{HMMHIDSevent}, the authors propose a Hidden Markov Models-based anomaly correlation approach for detecting suspicious activities in EVCSs. Their system used system logs and monitoring data to identify temporal patterns indicative of attacks. Similar to other HIDS approaches, they also worked with only benign data and simulated attack scenarios.

In Warraich and Morsi \cite{warraich2023early}, the authors propose an early detection system for cyber-physical attacks on fast charging stations. Their approach is particularly notable for its focus on the physical aspects of charging infrastructure and consideration of Vehicle-to-Grid operations in microgrid environments. They demonstrated that monitoring the power consumption patterns and electrical characteristics of EVCS could provide early warning signs of attacks, even before they manifest themselves as obvious network anomalies. 

Sharma et al. \cite{sharma2025artificial} proposed an AI-augmented architecture that incorporates both host-level monitoring and grid-level anomaly detection. Their system uses a hierarchical approach to correlate events across multiple layers of the charging infrastructure, enabling more accurate detection of sophisticated attacks that might affect both the cyber and physical layers simultaneously. However, the implementation of the architecture proposed by these authors is a centralized AI-driven anomaly detection framework, rather than a true dual-layer hybrid IDS. In other words, their framework lacks separate NIDS and HIDS components. It does not have dedicated network traffic analysis components that monitor flow-level and packet-level features, nor independent host-based detectors that analyze system calls and hardware performance counters. Their detection approach relies primarily on power consumption patterns and charging session behavior using Gaussian-based thresholding. It can be argued that such a monolithic design cannot detect network-level reconnaissance attacks, protocol-layer exploits, or early-stage attack indicators; it operates reactively, only after attacks impact operations, rather than proactively identifying threats at both the network and host levels.

More recently, host-level detection for EVCSs has gained momentum with the release of the CICEVSE2024 dataset. Alongside the dataset, Buedi et al.~\cite{CICEVSE2024} provide baseline machine learning results, reporting a weighted accuracy of 78.87\% for five-class classification of HPC events. Li et al. \cite{li2025multi} propose a multi-view graph contrastive representative learning approach that represents HPC logs from the EVSE as graph-structured data, achieving a weighted F1-score of 97.2\% on fine-grained multiclass classification. However, the approach leaves open questions in terms of real-world deployment and generalization ability. Al-E'mari et al.~\cite{al2025forensic} propose a forensic analysis framework that applies machine learning models to the host events (kernel and HPC) data of CICEVSE2024, achieving 98.81\% accuracy on a coarse-grained four-class problem (benign, cryptojacking, DoS, and reconnaissance). Rahal et al. \cite{rahal2025fuse} fuse network traffic and kernel events in a federated learning framework, reaching 98.91\% accuracy. However, their evaluation is limited to three coarse classes (benign, DoS, and reconnaissance) and covers neither host-based attacks such as cryptojacking and backdoor, nor FDIAs, nor the power consumption data source.

These existing approaches highlight a critical gap: earlier NIDS approaches for EVCS use generalized network traffic datasets, which do not reflect the realities in EVCS environments, while earlier HIDS approaches use EVCS-specific data but rely on simulated attacks rather than actual attack data. Most of the existing approaches also focus exclusively on either network or host-based detection, missing the potential benefits of a combined approach. Recent works such as Benfarhat et al. \cite{benfarhat2025advanced} use EVCS-specific datasets with real attack data, however, they focus mainly on network-level detection using advanced deep learning techniques. Similarly, Sharma et al. \cite{sharma2025artificial} and Warraich and Morsi \cite{warraich2023early} consider both cyber and physical aspects, but they do not fully implement a dual-layer detection architecture that can enable fine-grained network- and host-based monitoring. Even the most recent host-level and multimodal efforts presented in \cite{li2025multi, al2025forensic, rahal2025fuse} either classify only coarse attack categories or rely on host events data alone, and none of them exploits power consumption data, leaving FDIA detection unaddressed.

To overcome these gaps, we propose a dual-layered architecture that aligns with the cyber-physical structure of the EV charging ecosystem. Specifically, the network-based IDS is deployed at the CSMS, while multiple host-based IDSs are deployed at individual EVCSs to detect individual attacks at both layers.
Our proposed Hybrid IDS addresses three main research gaps in existing IDS for EVCSs: (1) the lack of complementary monitoring across cyber and physical layers, (2) limitations in data specificity and classification granularity, and (3) the absence of efficient data preprocessing and feature engineering pipelines that ensure optimal performance of the attack detection models. The recently released  CICEVSE2024 dataset \cite{CICEVSE2024} addresses data limitations by providing a comprehensive dataset specific to EVCS environments that includes both benign and actual attack data collected at the network and host layers. Hence, the Hybrid IDS proposed in this paper uses this dataset in training and validation of detection models for attacks at both network and host layers of the EVCS environment to ensure that it adequately reflects real-world conditions.

\section{Proposed Hybrid IDS Approach}
Building upon the analysis of existing approaches and their limitations, we present our proposed solution: a dual-layered hybrid intrusion detection system designed specifically for the unique requirements of intrusion detection in EVCS environments.

\subsection{Conceptual Architecture}
\begin{figure}[!t]
    \centering
    \includegraphics[width=0.9\linewidth]{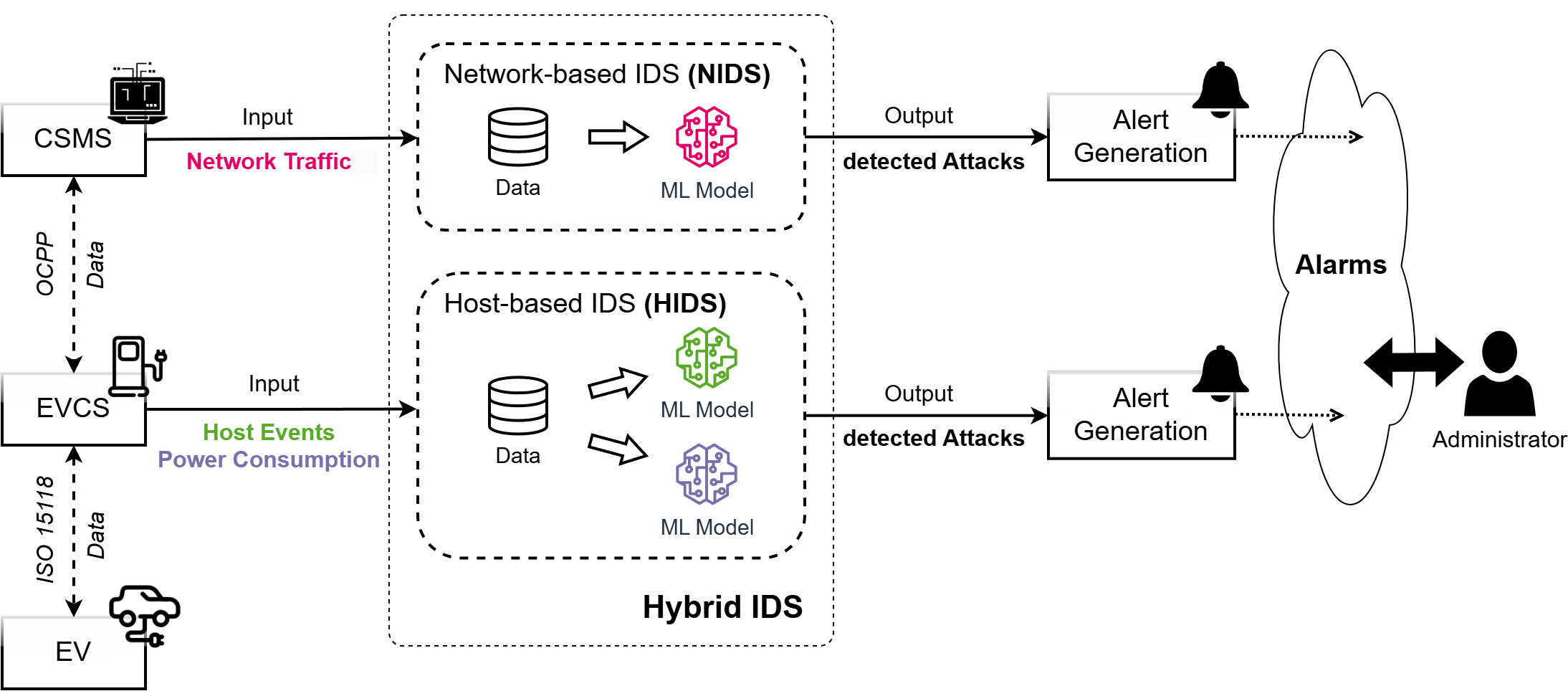}
    \caption{Conceptual architecture of the proposed Hybrid IDS.}
    \label{fig:hybrid_ids}
\end{figure}

As illustrated in Fig. \ref{fig:hybrid_ids}, the Hybrid IDS architecture consists of two main components: the NIDS and the HIDS components. The NIDS component captures network traffic from the CSMS while the HIDS component collects host event logs and power consumption data from individual EVCSs. The collected data are preprocessed to remove missing values. In the next step, the preprocessed data are analyzed using machine learning classifiers, which identify each sample as benign or a specific attack type. Upon detection of an attack, an alert is generated and transmitted to the network administrator, who can then respond accordingly. The proposed Hybrid IDS operates at the decision level: the NIDS and HIDS independently classify events and generate alerts, and an event is treated as malicious if either component detects anomalous behavior.

This coordinated approach enables comprehensive detection at both the network and host levels, effectively addressing threats towards the cyber-physical system. This is particularly important, as attacks such as denial of service attacks at the network layer can result in abnormal activities at the host level, for example, interruption of charging and ultimately power flow to the EV. Hence, by ensuring simultaneous attack detection at both levels, the proposed framework helps to reduce the chance of attacks going undetected.

\subsection{Dataset Description}

The CICEVSE2024 dataset \cite{CICEVSE2024} is composed of power consumption data, network traffic, and host event logs of EVCSs in both benign and attack scenarios. In the lab setup used to generate the data, there were two Electric Vehicle Supply Equipment (EVSE) units, serving as the physical interface within an EVCS: EVSE-A, a real EV charger, and EVSE-B, an EVSE emulated on a Raspberry Pi. The dataset was collected in different scenarios: benign, host-based attacks, and network-based attacks, each in both idle and charging states generating a total of six scenarios.

A total of 16 attacks were executed, of which cryptojacking and backdoor are host-based attacks, and the remaining 14 are network-based attacks, including seven types of reconnaissance attacks and seven types of DoS attacks. In addition to these sixteen attacks, FDIA scenarios were simulated and injected into both the network traffic and power consumption data, as described in Section~\ref{experiments}. The three data sources (network traffic, host events, and power consumption) have different data sizes, features, and were collected for different attack scenarios, as summarized in Table~\ref{tab:data_sources}. The table lists the features of the dataset as published and the processed data used in this work.

\begin{table}[!t]
\caption{Overview of the CICEVSE2024 Dataset as Published and as Used in This Work}
\label{tab:data_sources}
\centering
\scriptsize
\setlength{\tabcolsep}{3pt}
\renewcommand{\arraystretch}{1.1}
\begin{tabular}{|p{2.7cm}|c|c|c|}
\hline
\textbf{Data Source} & \textbf{Published} & \textbf{Used} & \textbf{Classes} \\
 & \textbf{(samples/features)} & \textbf{(samples/features)} & \textbf{(used)} \\
\hline
Network traffic flow-level, EVSE-A & 547,854 / 86 & 547,854 / 67 & 13 \\
\hline
Network traffic packet-level, from EVSE-A $^{a}$ & -- & 1,309,252 / 28 & 16 \\
\hline
Host events, EVSE-B$^{b}$ & 8,468 / 911 & 12,499 / 551 & 16 \\
\hline
Power consumption, EVSE-B$^{a}$ & 115,298 / 5 & 115,298 / 9 & 8 \\
\hline
\end{tabular}
\begin{flushleft}
\footnotesize
Benign data is included within the considered classes. \\ 
$^{a}$Includes the simulated FDIA class (Section~\ref{experiments}). \\
$^{b}$Re-extracted from the raw event logs to recover precise timestamps (Section~\ref{experiments}). 
\\ The network traffic of EVSE-B (2,196,846 flows, 86 features) was not used, as it contains no benign traffic.
\end{flushleft}
\end{table}

\subsubsection{Network Traffic Data}
The CICEVSE2024 dataset contains network traffic data from both EVSE-A and EVSE-B. The data was captured using a network topology where the Electric Vehicle Communication Controller (EVCC) communicates bidirectionally with EVSE-B through a switch using the ISO 15118 protocol, while EVSE-A and EVSE-B use the OCPP protocol to communicate with remote and local CSMS, respectively.

Two types of network traffic data were collected: flow-level data and packet-level data. Flow-level data, extracted using the \texttt{NFStream} Python library, provides features such as flow duration and the number of packets within each flow, whereas packet-level data offers more detailed information about individual packets, including protocol types, packet sizes, and header information.

\subsubsection{Host Events Data}

In CICEVSE2024, host events data were collected only from EVSE-B. Event logs of Hardware Performance Counters (HPC) and kernel events were collected under six different scenarios (benign, network attack, host-based attack each in idle and charging state).

\subsubsection{Power Consumption Data}
Power consumption data were likewise collected only from EVSE-B. Voltage, current, and power measurements were recorded by an Inter-Integrated Circuit (I2C) wattmeter monitoring the power supplied to EVSE-B under both benign and attack scenarios, yielding five features. Although the data was sampled at a one-second rate, the timestamps are only minute-level with occasional skipped seconds, and the power consumption data covers fewer attack scenarios than the other two data sources.

\subsection{Attack Detection Framework}
The goal of the proposed Hybrid IDS is to identify and distinguish between various network- and host-based attacks. For network-based attacks, the system is trained to identify normal traffic, Reconnaissance (Recon) attacks, Denial of Service (DoS) attacks, and False Data Injection Attacks (FDIAs). For host-based attacks, the system is trained to detect normal events, backdoor attacks, and cryptojacking.

Formally, let $X = \{x_1, x_2, ..., x_n\}$ denote a set of input features extracted from network traffic, host events, or power consumption data. The classification task performed by the proposed Hybrid IDS can then be defined as finding a function $f$ that maps these features to a set of attack classes $C = \{c_1, c_2, ..., c_m\}$:

\begin{equation}
f: X \rightarrow C
\end{equation}

where $c_1$ typically represents benign traffic/events and $c_2$ through $c_m$ represent different attack types. The objective of the Hybrid IDS is to minimize the classification error:

\begin{equation}
\min_{f} \sum_{i=1}^{N} \mathbb{I}(f(x_i) \neq y_i)
\end{equation}

where $y_i$ is the true label of sample $i$, $N$ is the total number of samples, and $\mathbb{I}$ is the indicator function.
We employ the following tree-based ensemble ML algorithms in the classification process: Random Forest (RF), XGBoost, Light Gradient Boosting Machine (LightGBM), and Category Boosting (CatBoost). These algorithms were chosen for their good performance as reported in prior intrusion detection research and their ability to handle the types of features present in the chosen dataset.

\section{Experimental Setup} \label{experiments}

In this section, we discuss the procedure used to validate the performance of the proposed Hybrid IDS. We first discuss the data processing and feature engineering approach we employed, after which we present the training and validation of the ML classifiers.
\subsection{Data Processing and Feature Engineering}
The data processing and feature engineering workflow for each data source follows a systematic process comprising three main phases: (1) data reorganization and integration, (2) preprocessing and feature extraction, and (3) feature selection and transformation. Each data type required specific handling techniques to address its unique characteristics and challenges.

\subsubsection{Network Traffic Data Processing}
For network traffic data, we focused on data relevant to EVSE-A due to its completeness and representativeness. Two complementary approaches were implemented:

\textit{Flow-based Analysis:} The original flow-level datasets were integrated into a unified dataset with standardized labels. The original CICEVSE2024 dataset contains 86 features extracted using \texttt{NFStream}, out of which we used 67 after eliminating non-numeric features, columns with missing data or zero variance. Some of the features used include temporal attributes (e.g., flow duration, inter-arrival time), volumetric features (e.g., packet count, byte count), and protocol-specific metrics. Address information (e.g., IP, MAC, and OUI) was excluded from the dataset to prevent potential data leakage. This decision was made based on the dynamic nature of real-world network environments, which contrasts with the relatively static and limited number of devices in the simulation. Quantitative analysis revealed a significant class imbalance, with benign traffic constituting 0.015\% of all flows. To address this, we applied Synthetic Minority Over-sampling Technique (SMOTE) with a sampling strategy optimized to achieve a more balanced representation while avoiding overfitting. However, SMOTE either decreased the performance or did not affect the choice of the best classifier.

\textit{Packet-based Analysis:} Packet-level data was extracted from the original \texttt{pcap} files available within the dataset using the \texttt{Scapy} Python library. We processed the raw packet captures through a multi-stage pipeline: (1) packet extraction and parsing, (2) feature derivation, and (3) dimensionality optimization. The packets were organized into a structured format, with 38 distinct features extracted per packet. After performing feature engineering, 28 features were left, including TCP flags and payload characteristics.
To expand the detection scope of the NIDS, FDIAs were simulated within the packet-level network traffic dataset using a pattern-based manipulation technique based on attack signatures observed in similar IoT environments \cite{FDIA_network}. Specifically, half of the benign samples were randomly selected and modified to exhibit FDIA characteristics: for each original MAC address, a fake MAC address differing only in the last byte was generated to represent an attacker device, and the \texttt{ip\_ttl} value was reduced by one to mimic the detouring behavior of the FDIA pattern. Modifying half of the benign samples ensured a balanced representation of benign and attack scenarios. This simulation was limited to the packet-level data for two reasons. First, the combined flow-level dataset contains only 82 benign samples, so halving it would create another minority class and worsen the existing imbalance. Second, the TTL feature required to simulate the FDIA behavior is not available in the flow-level data, making a MAC-address-only modification insufficient for an accurate simulation.

\subsubsection{Host Events Data Processing}
In the case of host events data, we constructed a multi-dimensional feature space from heterogeneous log sources. The preprocessing pipeline combined Hardware Performance Counters (HPC) data, capturing microarchitectural events (e.g., cache misses, branch prediction outcomes) with kernel-level system events (e.g., system calls, I/O operations). Because the published host events file provides only relative time offsets, the host events data were re-extracted from the raw event logs to recover precise timestamps, yielding 12,499 records; after preprocessing and the removal of zero-variance features, 551 features were used.

\subsubsection{Power Consumption Data Processing}
Power consumption data required specialized processing to extract meaningful patterns from time-series measurements owing to the low training and evaluation performances initially.
We developed a four-stage feature extraction framework consisting of: (1) Data Cleaning: removal of duplicate rows and handling of missing values through cubic spline interpolation; (2) Feature Engineering: derivation of new features using a rolling window technique; (3) Scaling and Class Imbalance Management: application of scaling techniques like MinMaxScaler and StandardScaler, along with handling class imbalance through methods such as class\_weight and SMOTE, which ultimately demonstrated limited success in improving performance; (4) Final Preprocessing: selection of StandardScaler as the optimal scaling method based on its effectiveness in addressing low standard deviation issues within key features. Additionally, FDIAs were simulated in the power consumption data by increasing or decreasing the \texttt{power\_mW} values of a continuous segment of benign data by 20\%, with the corresponding \texttt{current\_mA} values adjusted proportionally to preserve consistency under the relation $P = V \times I$, assuming constant voltage. This perturbation level was selected based on prior studies showing that sudden load changes can significantly affect grid stability \cite{ImpactOfEVBotnetsOnPowerGrid2019, Morrison2018}.

\subsection{ML Classifier Training and Validation} \label{subsec:valid_method}

For each data source, we implemented and compared multiple machine learning classifiers: Random Forest (RF), XGBoost, LightGBM, and CatBoost, with a Decision Tree (DT) classifier additionally evaluated on the power consumption data. Kernel-based Support Vector Machines (SVMs) and single DTs were considered for the network traffic and host events data but were excluded owing to their poor scalability and tendency to overfit on these large, high-dimensional datasets, respectively.

Due to their large size (547,854 flow-level and 1,309,252 packet-level records), the network traffic datasets were split into training (80\%), validation (10\%), and testing (10\%) sets. The host events (12,499 records) and power consumption (115,298 records) datasets were split into training (60\%), validation (20\%), and testing (20\%) sets. For the network traffic and host events data, stratified random splitting was applied to preserve class proportions across all subsets, which is important owing to the strong class imbalance in the host events data (refer to Fig.~\ref{fig:cd_hostdata}). For the power consumption data, a combined chronological and stratified split was used to preserve the temporal order introduced by the rolling-window features while approximately maintaining class proportions. In both the packet-level and power consumption datasets, the FDIA simulation was performed before the data were split. All preprocessing steps were implemented with the \texttt{scikit-learn} library, and the classifiers were taken from the \texttt{scikit-learn}, \texttt{XGBoost}, \texttt{LightGBM}, and \texttt{CatBoost} libraries.

Hyperparameter optimization was performed using Bayesian optimization with tree-structured Parzen estimators to find the optimal configuration for each classifier. The primary evaluation metrics included accuracy, precision, recall, F1-score, and G-mean.

For a multiclass classification problem with $k$ classes, these metrics are defined as follows:

\begin{align}
\text{Accuracy} &= \frac{\sum_{i=1}^{k} TP_i}{N} \\
\text{Precision}_i &= \frac{TP_i}{TP_i + FP_i} \\
\text{Recall}_i &= \frac{TP_i}{TP_i + FN_i} \\
\text{F1-score}_i &= 2 \times \frac{\text{Precision}_i \times \text{Recall}_i}{\text{Precision}_i + \text{Recall}_i}
\end{align}

where $TP_i$, $FP_i$, and $FN_i$ represent true positives, false positives, and false negatives for class $i$, respectively, and $N$ is the total number of samples. The G-mean, which provides a balanced measure of performance across all classes, is calculated as:

\begin{equation}
\text{G-mean} = \sqrt[k]{\prod_{i=1}^{k} \text{Recall}_i}
\end{equation}

where $k$ is the number of classes. This metric is particularly useful for imbalanced datasets, as it is sensitive to performance on minority classes. For multiclass problems, we report the macro-average of these metrics across all classes owing to the class imbalance, ensuring that a large, well-classified class does not dominate the overall metric.

\section{Results}
After conducting validation experiments based on the data processing pipelines and model training methodology described in Section \ref{experiments}, we present the results of the experiments conducted and analyze the performance of our hybrid approach in comparison with the methods proposed in related works. All results are reported on the held-out test sets (10\% of the network traffic data and 20\% of the host events and power consumption data), following the splitting methodology described in Section~\ref{subsec:valid_method}. 

\subsection{Network-based Detection Results}
The performance of different classifiers for network-based detection varies depending on whether flow-level or packet-level data is used.

During validation, we found that flow-level network traffic data had better performance and lower prediction time than packet-level data. Based on testing with flow-level network traffic data, the XGBoost classifier achieved the highest performance for the NIDS, with an accuracy of 99.99\%, precision of 99.98\%, recall of 99.96\%, F1-score of 99.97\%, and G-mean of 99.98\%, with a prediction time of 0.706~$\mu$s per sample.
The testing results demonstrated only five misclassifications out of 54,784 samples, all of which involved misclassifying one attack type as another, while maintaining perfect distinction between normal and attack traffic. However, benign traffic is extremely scarce in the flow-level data (82 samples, i.e., 0.015\% of all flows, of which only eight fall into the testing set). The perfect benign--attack separation should therefore be interpreted with this class imbalance in mind. The packet-level analysis described in Section~\ref{experiments}, in which benign samples are far more numerous, was conducted in part to mitigate this limitation.

\begin{table*}[!b]
\caption{Performance Comparison with Existing Approaches}
\label{tab:comparison}
\centering
\setlength{\tabcolsep}{4pt}
\renewcommand{\arraystretch}{0.9}
\begin{tabular}{|l|c|c|c|c|c|c|c|}
\hline
\textbf{Approach} & \textbf{Accuracy (\%)} & \textbf{Precision (\%)} & \textbf{Recall (\%)} & \textbf{F1-score (\%)} & \textbf{Classes} & \textbf{Dataset} & \textbf{Detection Type} \\
\hline
\multicolumn{8}{|c|}{\textbf{Network-based Detection Systems}} \\
\hline
ElKashlan et al. \cite{MLIDSEVCS} & 99.20 & 98.50 & 99.00 & 98.70 & 5 & IoT23 & Network-based \\
\hline
Basnet et al. \cite{DLIDSEVCS} & - & 100.00 & 99.80 & 99.80 & 5 & CICIDS2018 & Network-based \\
\hline
Buedi (NADM) \cite{multi-stageIDS} & 98.00 & 99.00 & 98.00 & 98.00 & 2 & EVCS-specific & Network-based \\
\hline
FL-based ADS \cite{FLADS} & 96.97 & - & - & 97.40 & 2 & CICEVSE2024 & Network-based \\
\hline
Jiang et al. \cite{Jiang_CICEVESE_KD} & 92.48 & 93.13 & - & 92.20 & 15 & CICEVSE2024 & Network-based \\
\hline
Benfarhat et al. \cite{benfarhat2025advanced} & 93.90 & 94.10 & 93.30 & 93.90 & 17 & CICEVSE2024 & Network-based \\
\hline
\textbf{Proposed NIDS} & \textbf{99.99} & \textbf{99.98} & \textbf{99.96} & \textbf{99.97} & \textbf{13} & \textbf{CICEVSE2024} & \textbf{Network-based} \\
\hline
\multicolumn{8}{|c|}{\textbf{Host-based and Multimodal Detection Systems}} \\
\hline
Buedi et al. \cite{CICEVSE2024}$^{a}$ & 78.87 & 77.83 & 78.87 & 78.22 & 5 & CICEVSE2024 & Host-based \\
\hline
Al-E'mari et al. \cite{al2025forensic}$^{b}$ & 98.81 & 98.81 & 98.81 & 98.81 & 4 & CICEVSE2024 & Host-based \\
\hline
Rahal et al. \cite{rahal2025fuse}$^{c}$ & 98.91 & - & - & 98.90 & 3 & CICEVSE2024 & Multimodal \\
\hline
\rowcolor[HTML]{EFEFEF}
Proposed HIDS (host events)$^{d}$ & 96.60 & 85.25 & 85.58 & 85.31 & 16 & CICEVSE2024 & Host-based \\
\hline
\rowcolor[HTML]{EFEFEF}
Proposed HIDS (power consumption)$^{d}$ & 70.34 & 73.78 & 71.40 & 70.57 & 8 & CICEVSE2024 & Host-based \\
\hline
\rowcolor[HTML]{EFEFEF}
Proposed HIDS (combined)$^{d}$ & 83.47 & 82.18 & 82.00 & 81.47 & 17 & CICEVSE2024 & Host-based \\
\hline
\end{tabular}
\begin{flushleft}
\footnotesize
Classes: number of classification labels including benign; `-' indicates the value is not reported in the cited work.
$^{a}$Random Forest on HPC events; weighted averages over five classes (benign, cryptojacking, backdoor, reconnaissance, DoS).
$^{b}$XGBoost/LightGBM on host events (kernel and HPC); four classes (benign, cryptojacking, DoS, reconnaissance).
$^{c}$Federated fusion (10 clients) of network traffic and kernel events; three classes (benign, DoS, reconnaissance); overall precision and recall not reported.
$^{d}$Macro-averaged over fine-grained attack classes: 16 classes (host events), 8 classes (power consumption), 17 classes combined.
\end{flushleft}
\end{table*}

\begin{table*}[!t]
\caption{Classification Report of the HIDS on Host-based Intrusion Detection}
\label{tab:classreport}
\centering
\setlength{\tabcolsep}{3pt}
\renewcommand{\arraystretch}{1.05}
\begin{tabular}{|l|l|ccc|ccc|ccc|}
\hline
\multicolumn{2}{|c|}{\multirow{2}{*}{\textbf{Attack Class}}} & \multicolumn{3}{c|}{\textbf{Host Events Part}} & \multicolumn{3}{c|}{\textbf{Power Consumption Part}} & \multicolumn{3}{c|}{\textbf{HIDS}} \\
\cline{3-11}
\multicolumn{2}{|c|}{} & \textbf{Prec.} & \textbf{Rec.} & \textbf{F1} & \textbf{Prec.} & \textbf{Rec.} & \textbf{F1} & \textbf{Prec.} & \textbf{Rec.} & \textbf{F1} \\
\hline
\multicolumn{2}{|l|}{Benign} & 1.00 & 1.00 & 1.00 & 0.99 & 0.92 & 0.95 & 0.995 & 0.96 & 0.975 \\
\hline
\multirow{7}{*}{\rotatebox{90}{Recon}}
 & TCP Port Scan & 0.75 & 0.90 & 0.82 & -- & -- & -- & 0.75 & 0.90 & 0.82 \\
 & Service Version Detection & 0.50 & 0.45 & 0.47 & -- & -- & -- & 0.50 & 0.45 & 0.47 \\
 & OS Fingerprinting & 0.84 & 0.86 & 0.85 & -- & -- & -- & 0.84 & 0.86 & 0.85 \\
 & Aggressive Scan & 0.82 & 0.75 & 0.78 & -- & -- & -- & 0.82 & 0.75 & 0.78 \\
 & SYN Stealth Scan & 0.63 & 0.63 & 0.63 & 0.20 & 0.02 & 0.04 & 0.415 & 0.325 & 0.335 \\
 & Vulnerability Scan & 0.61 & 0.56 & 0.59 & 0.56 & 0.31 & 0.40 & 0.585 & 0.435 & 0.495 \\
 & Slowloris Scan & -- & -- & -- & -- & -- & -- & -- & -- & -- \\
\hline
\multirow{7}{*}{\rotatebox{90}{DoS}}
 & UDP Flood & 0.96 & 1.00 & 0.98 & -- & -- & -- & 0.96 & 1.00 & 0.98 \\
 & ICMP Flood & 0.72 & 0.84 & 0.78 & -- & -- & -- & 0.72 & 0.84 & 0.78 \\
 & PSHACK Flood & 1.00 & 1.00 & 1.00 & -- & -- & -- & 1.00 & 1.00 & 1.00 \\
 & ICMP Fragmentation & 0.85 & 0.78 & 0.81 & -- & -- & -- & 0.85 & 0.78 & 0.81 \\
 & TCP Flood & 1.00 & 0.96 & 0.98 & 0.43 & 0.77 & 0.55 & 0.715 & 0.865 & 0.765 \\
 & SYN Flood & 1.00 & 1.00 & 1.00 & 0.92 & 0.85 & 0.88 & 0.96 & 0.925 & 0.94 \\
 & SynonymousIP Flood & 0.96 & 0.96 & 0.96 & -- & -- & -- & 0.96 & 0.96 & 0.96 \\
\hline
\multicolumn{2}{|l|}{Cryptojacking} & 1.00 & 1.00 & 1.00 & 0.96 & 0.92 & 0.94 & 0.98 & 0.96 & 0.97 \\
\hline
\multicolumn{2}{|l|}{Backdoor} & 1.00 & 1.00 & 1.00 & 0.86 & 1.00 & 0.92 & 0.93 & 1.00 & 0.96 \\
\hline
\multicolumn{2}{|l|}{FDIA} & -- & -- & -- & 0.99 & 0.93 & 0.96 & 0.99 & 0.93 & 0.96 \\
\hline
\end{tabular}
\begin{flushleft}
\footnotesize
`--' indicates that the attack class is not present in the corresponding data source; in particular, the Slowloris Scan class was excluded from the host events data owing to an insufficient number of samples (eight) and does not occur in the power consumption data.
\end{flushleft}
\end{table*}

\begin{figure}[!t]
    \centering
    \includegraphics[width=0.8\linewidth]{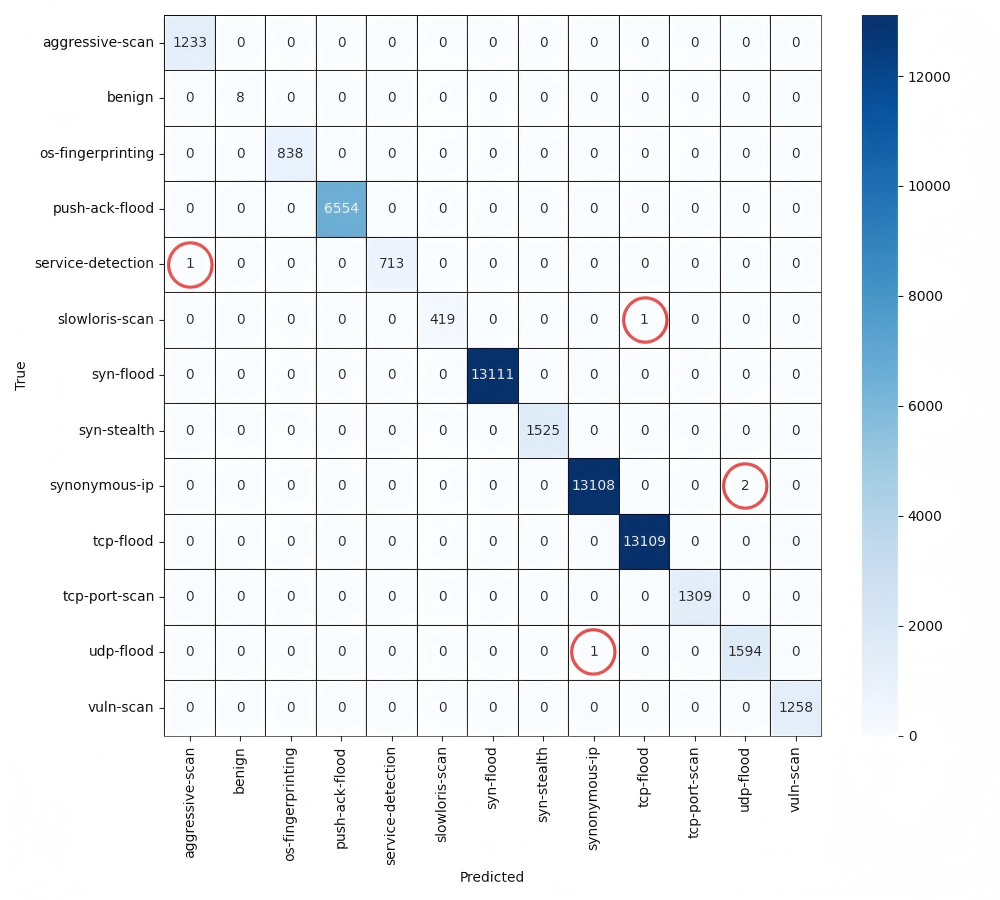}
    \caption{Confusion matrix for network-based detection based on flow-level data}
    \label{fig:cm_nids}
\end{figure}

The confusion matrix shown in Fig.~\ref{fig:cm_nids} was computed with \texttt{scikit-learn} from the predicted and true class labels of the flow-level NIDS test set (54,784 samples). It shows that the NIDS effectively distinguishes between different attack types, with some minor confusion between similar attack variants.

\subsection{Host-based Detection Results}
\begin{figure}[!t]
    \centering
    \includegraphics[width=0.8\linewidth]{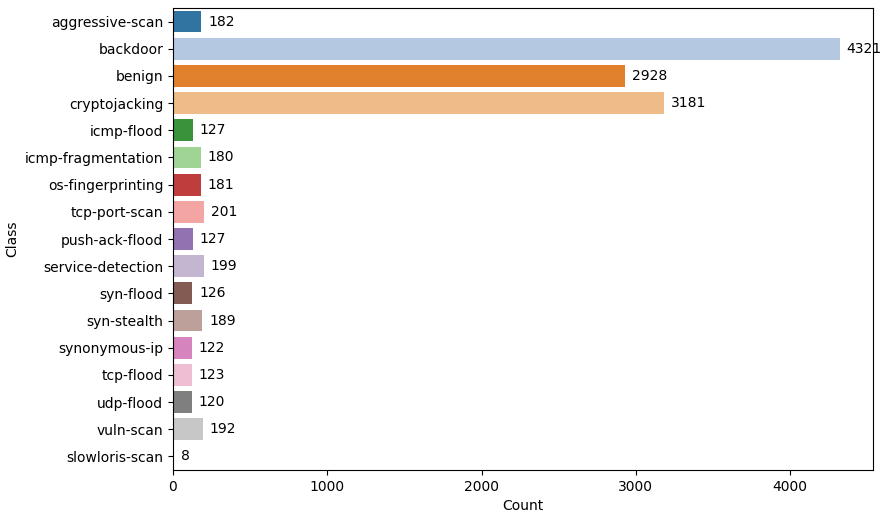}
    \caption{Class distribution of the host events data}
    \label{fig:cd_hostdata}
\end{figure}

For host events data, the XGBoost classifier achieved the best performance, with an accuracy of 96.60\%, precision of 85.25\%, recall of 85.58\%, F1-score of 85.31\%, and G-mean of 92.41\%. The discrepancy between the high accuracy and the lower macro F1-score is due to class imbalance in the host events dataset, whose class distribution is shown in Fig.~\ref{fig:cd_hostdata}. As seen in the figure, benign samples and host-based attacks significantly outnumber the network-based attack classes. The model classifies these majority classes well, which keeps the overall accuracy high, while its performance on the underrepresented classes is weaker. Since accuracy is dominated by the most frequent classes, it may remain high even when minority-class instances are misclassified. In contrast, the macro F1-score, which reflects the balance between precision and recall across all classes, is more sensitive to such errors and therefore better reveals the impact of class imbalance.

The host events model showed particularly strong performance in detecting benign activities, cryptojacking, backdoor attacks, and certain DoS attacks like PSHACK Flood, SYN Flood, and SynonymousIP Flood, all with F1-scores of 0.96 or higher.

For power consumption data, the LightGBM classifier achieved an accuracy of 70.34\%, with precision of 73.78\%, recall of 71.40\%, and F1-score of 70.57\%. While this performance is lower than the host events model, it excelled at detecting specific attack types, particularly FDIAs (F1-score of 0.96), cryptojacking (F1-score of 0.94), and backdoor attacks (F1-score of 0.92).

More details about the per-class performance of the proposed HIDS are presented in Section~\ref{subsec:per_class_report}.

\subsection{Comparison with Existing Works}

The performance comparison with existing approaches is summarized in Table \ref{tab:comparison}. While direct comparison is challenging due to differences in datasets and evaluation methods, our approach shows competitive or superior performance across different attack categories. Compared to existing approaches that focus solely on either network or host-based detection, our Hybrid IDS offers several advantages: (1) More comprehensive coverage of attack vectors by monitoring both network traffic and host activities; (2) Higher detection accuracy for specific attack types by utilizing data specific to EVCS environments; (3) Ability to detect a wider range of attacks, including those that might evade single-source detection; and (4) Better context awareness by correlating information from multiple sources.

A notable recent contribution in this field is the advanced Temporal Convolutional Network framework proposed by Benfarhat et al. \cite{benfarhat2025advanced}. Their MRG-ID-SA-TCN model achieves impressive performance on the same CICEVSE2024 dataset, with an overall accuracy of 93.9\%, precision of 94.1\%, recall of 93.3\%, and F1-score of 93.9\% for the 17-class problem (including benign and 16 attack types). Their approach excels at detecting high-priority attacks like ICMP flood, SYN flood, and Backdoor attacks with perfect detection in most cases. However, it shows some limitations with less common attacks such as Vulnerability Scan and Service Version Detection. In contrast, our hybrid approach demonstrates better performance for these specialized attack types.

On the host-based side, Table \ref{tab:comparison} compares the proposed HIDS with recent works evaluated on the same CICEVSE2024 host data. Compared with the baseline of Buedi et al.
\cite{CICEVSE2024}, which reports a weighted accuracy of 78.87\% for five-class classification of HPC events, our host events model achieves 96.60\% accuracy on a considerably finer-grained 16-class problem. Li et al. \cite{li2025multi} also perform fine-grained classification of HPC events and report a weighted F1-score of 97.2\%. However, weighted averages computed on their benign-dominated data are not directly comparable to the macro-averaged scores reported in this work, which weight all classes equally. Their results are therefore not included in Table \ref{tab:comparison}. Al-E'mari et al.~\cite{al2025forensic}
and Rahal et al.~\cite{rahal2025fuse} report higher headline accuracies (98.81\% and 98.91\%, respectively). However, both address much coarser tasks with only four and three attack categories, respectively, and neither utilizes the power consumption data source. In particular, the multimodal approach of Rahal et al. covers neither host-based attacks (cryptojacking, backdoor) nor FDIAs. In contrast, the proposed HIDS performs fine-grained classification across 17 attack classes and leverages power consumption data to detect FDIAs (F1-score of 0.96), which none of the compared approaches cover.

Overall, the evaluation demonstrates the effectiveness of the proposed hybrid approach in detecting a wide variety of attacks against EV charging infrastructure. The proposed NIDS achieves state-of-the-art performance on CICEVSE2024, while the proposed HIDS provides the most fine-grained host-based attack classification reported on this dataset to date and uniquely covers FDIAs through power consumption data, complementing detection capabilities that single-source approaches do not provide.

\subsection{Per-Class Detection Performance on Host-based Intrusion Detection} \label{subsec:per_class_report}

Table \ref{tab:classreport} provides a detailed per-class breakdown of the HIDS performance across its two host-based data sources, reporting the per-class precision, recall, and F1-score obtained from the host events and power consumption models individually alongside their combined HIDS performance. Since an alert is generated whenever either constituent model detects an attack, the per-class performance of the HIDS is computed as the macro-average of the two models for each class. The results show that host events data drives the detection of most host-relevant attacks, while power consumption data contributes complementary coverage for FDIAs and several DoS variants. Reconnaissance and certain DoS attacks remain challenging for both sources, as they lack distinctive host-level or power-consumption signatures. Given the near-perfect detection of these network-based attacks by the NIDS, this confirms the complementary nature of the two components, each optimized for distinct attack types. This complementarity further solidifies the effectiveness of the Hybrid IDS in enhancing the security of the EV charging ecosystem. 

\section{Conclusion and Future Work}
In this paper, we proposed a Hybrid IDS for EVCSs that combines network-based and host-based detection to provide comprehensive security monitoring across both cyber and physical layers. Leveraging the CICEVSE2024 dataset, which contains network traffic, host events, and power consumption data from actual EVCS environments, our system demonstrated high detection accuracy for various attack types.

The evaluation results show that the NIDS component of the proposed Hybrid IDS achieves 99.99\% accuracy for network-based attacks, while the HIDS component achieves 83.47\% overall accuracy for host-based intrusion detection. The HIDS demonstrated exceptional performance for specific attack types, particularly FDIAs (F1-score of 0.96), backdoor (F1-score of 0.96), and cryptojacking (F1-score of 0.97). These results demonstrate the effectiveness of our approach in detecting and classifying a wide range of attacks targeting EVCSs. Using the dual-layer integration, this approach facilitates intrusion detection at the network level and the host level.

While recent works such as Benfarhat et al. \cite{benfarhat2025advanced} have achieved good performance (93.9\% accuracy) using advanced deep learning approaches like temporal convolutional networks, our hybrid approach provides more comprehensive coverage by monitoring both network traffic and host activities. Similarly, the multi-view graph contrastive learning approach proposed by Li et al. \cite{li2025multi} shows promise for host-level detection based on HPC events, but it considers neither network traffic nor power consumption data, and thus lacks the dual-layer coverage that our hybrid approach provides. Likewise, the most recent host-level and multimodal efforts on the same dataset
\cite{al2025forensic, rahal2025fuse} classify only three to four coarse attack categories, whereas the proposed system provides fine-grained per-attack classification and is the only approach among these that exploits power consumption data.

The early detection framework for cyber-physical attacks presented by Warraich and Morsi \cite{warraich2023early} and the AI-augmented architecture proposed by Sharma et al. \cite{sharma2025artificial} offer valuable perspectives on microgrid-specific and architectural considerations, respectively, which could be integrated into future iterations of our system.

Despite these promising results, some limitations must be acknowledged. First, aggregating the outputs of the three data sources is currently infeasible. Benign network traffic is only available for EVSE-A, while host events and power consumption data originate from EVSE-B; moreover, timestamp irregularities in the power consumption data prevent even timestamp-based correlation of the HIDS detection outputs. Second, the high dimensionality of the host events features poses challenges for deployment on resource-constrained EVCSs, as reducing these features significantly degrades detection performance; deploying the HIDS at the CSMS or within other CPO infrastructure may therefore be more practical. Third, as a supervised approach, the proposed system is limited to detecting known attacks whose patterns are present in the training data.

There are several directions for future work. First, real-time detection capabilities could be enhanced by optimizing feature extraction and classification processes. Second, the correlation between different data sources could be further explored to improve the detection accuracy and reduce false positives. Third, automated response mechanisms could be integrated to provide immediate mitigation of detected attacks. Finally, the framework could be extended to incorporate data from additional sources, such as EV charging behavior data, to enhance the detection of sophisticated attacks targeting the entire EV charging ecosystem.


\bibliographystyle{IEEEtran}
\bibliography{bibliography}

\end{document}